\documentclass[12pt]{iopart}
\usepackage[T1]{fontenc}
\usepackage[latin1]{inputenc}
\usepackage{subfigure}
\usepackage{floatflt}
\usepackage{graphicx}
\usepackage{setspace}

\makeatletter

\usepackage{iopams}
\makeatother
\begin{document}

\title{Full Jet Reconstruction in d+Au and p+p Collisions at RHIC}

\author{Thomas Henry$^{\dagger}$ for the STAR 
Collaboration\footnote{For the full author list and acknowledgements see                               
Appendix "Collaborations" in this volume.}}

\address{$^{\dagger}$Texas A\&M University, College Station, TX 77843, USA\\
thenry@physics.tamu.edu}


\begin{abstract}
The STAR detector is well suited for investigating jet
production at RHIC. It has a large acceptance for both charged particles
and electromagnetic radiation, so that reconstructed jets contain a large
fraction of the particles descending from an initial hard scattered
parton. Inclusive jets can be used to measure $j_T$, which characterizes the
transverse jet shape.  Di-jet angular distributions in d+Au 
and p+p collisions at RHIC provide a determination of intrinsic and nuclear 
$k_{T}$. 
The most recent results from this analysis based on year 2003 d+Au and p+p 
data are presented and compared to results of a di-hadron correlation 
study.
\end{abstract}


\section{Introduction}

Jet studies are of great importance for understanding Au+Au
collisions at RHIC and the properties of the high density matter which 
result from them.  It is essential to understand how the 
medium does or does not modify jet yields and jet shapes \cite{Medium Mod}.  
The results from d+Au and p+p collisions provide the 
baseline for medium modification studies in Au+Au collisions.
For these simpler systems, jets can be examined
in great detail via full jet reconstruction.
In contrast, full jet reconstruction is not practical in 
Au+Au collisions, so the 
comparisons with Au+Au must be done using inclusive particle yields and
di-hadron and N-hadron correlations.

This study presents the first full jet reconstruction at RHIC and reports
$j_T$ distributions from p+p collisions
and $k_T$ measurements in p+p and d+Au collisions.  $j_T$ 
characterizes the
transverse shape of a jet.  $k_T$ measures the transverse momentum of a
parton within a nucleon or nucleus, and its growth from p+p to nuclear
collisions is believed to contribute to the Cronin effect \cite{cronin}.
Comparisons are also made to $j_T$ and $k_T$ results inferred from
di-hadron correlations.

The STAR detector is well suited for investigating jet
production at RHIC, due to the complete $\phi$ coverage and large $\eta$
coverage of the TPC and EMC.
The STAR Barrel EMC data used in this study
provide neutral energy including $\pi^{0}$ decay photons. 
The barrel EMC during the 2003 RHIC run measured neutral energy in the range 
$0<\eta<1$.
The 2400 towers each covered an angular area
of $\Delta\eta\times\Delta\phi=0.05\times0.05$. The barrel EMC was
read out in minimum bias events. It was also used to trigger on 
{}``high tower'' events, where one of the towers was
above a certain energy threshold, e.g. $E_{T}>2.5$ GeV.
This {}``high tower'' triggered event sample contains a much larger
fraction of jets than the minimum bias event sample.

This analysis uses two different algorithms that
reconstruct a jet by capturing the spray of fragmenting particles
in a geometric cone.  One centers the cone on the most energetic hadrons
in the event, while the other optimizes the cone direction to maximize
the included energy.
Care must be taken when reconstructing jets in d+Au 
collisions, to ensure that an algorithm is robust at high multiplicity. 

Measurement of jet energy requires corrections for charged particle energy
deposition in the EMC, for the finite efficiencies of the TPC and EMC, 
and for the unmeasured energy carried by long-lived neutral particles 
($n$, $K_L$, ...).
It is also possible for the jet reconstruction algorithm to miss soft 
particles.  The total energy scale correction applied is
about $1/0.8$ for minimum bias events, and $1/0.86$ for 
{}``high tower'' events, and is derived from Pythia Monte Carlo Simulations.

A sample of Pythia events processed by the STAR detector simulator was 
analyzed in the same fashion as the data. 
For jets with measured $E_{T} < 9$ GeV,
the Pythia simulations compare well with the data. Understanding
the higher $E_{T}$ jet distributions is still in progress, though
low to moderate $E_T$ jets dominate
average quantities such as $\left<E_{T}\right>$ and $\left<j_{T}\right>$. 

\section{Inclusive Jet Studies}
\begin{floatingfigure}[r]{0.50\columnwidth}%
\begin{spacing}{0.8}
\hspace*{-0.4in}
\includegraphics*[%
  width=0.50\columnwidth,
  keepaspectratio]{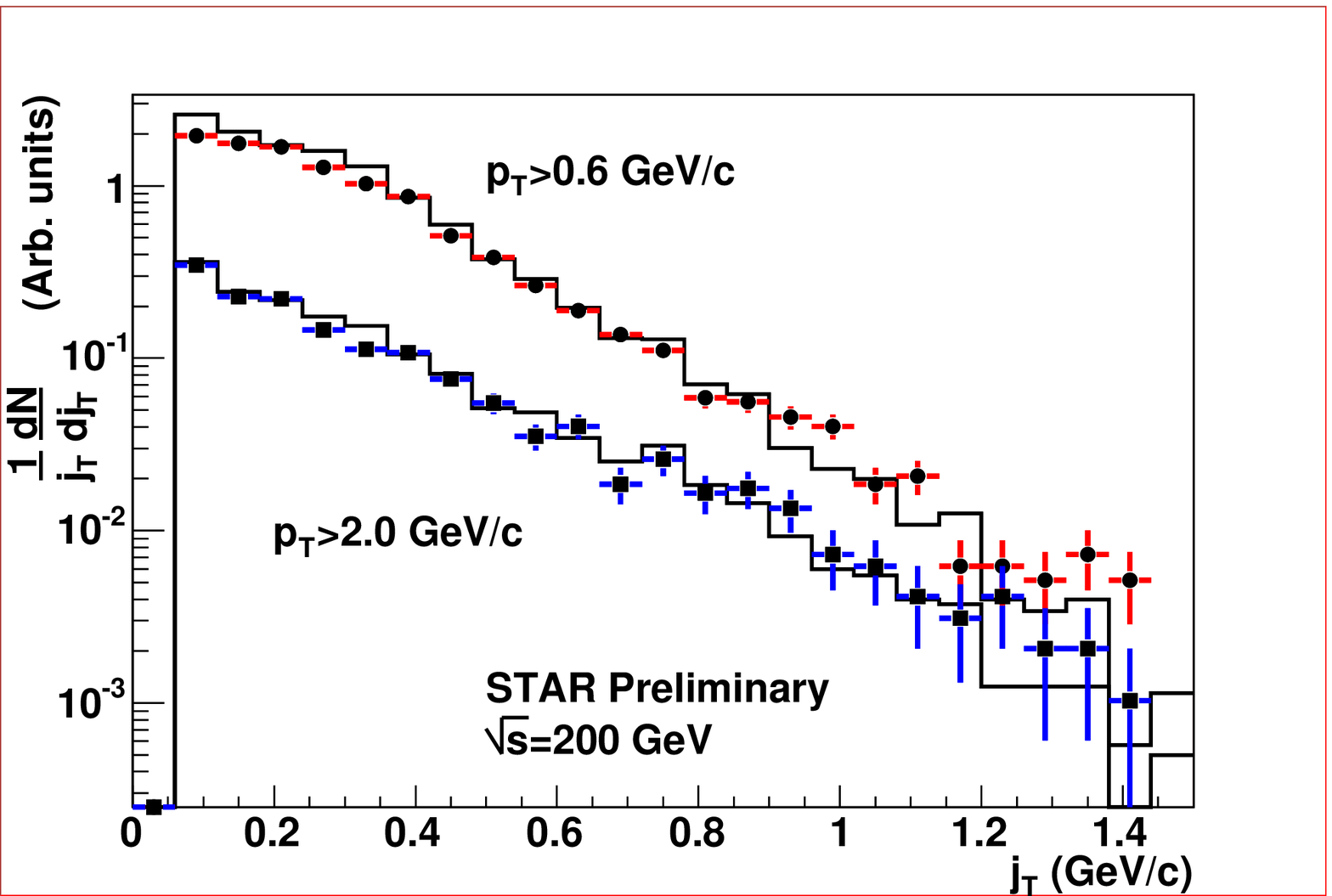}
\vspace*{-.25in}
\caption{p+p jet $j_T$}
\noindent \footnotesize Jet $j_T$ distributions for two hadron $p_T$ 
thresholds.
Histograms are Pythia Monte Carlo events processed by the STAR detector
simulator. The Pythia curve for hadron $p_{T}>2.0$ GeV/c has been arbitrarily
normalized for comparison purposes. \smallskip{}
\end{spacing}
\end{floatingfigure}%
The component of the hadron momentum perpendicular to the jet
thrust axis is given by: 

\vspace*{0.25in}
$j_{T}=\sqrt{p_h^{2}-\frac{\left(\vec{p_h}\cdot \vec{p_j}\right)^{2}}{p_j^{2}}},$
\vspace*{0.25in}

\noindent where $\vec{p_h}$ and $\vec{p_j}$ represent the momentum of the 
hadron and jet, respectively.  The formula shows that while 
$j_{T}$ is related to the fragmentation of the jet and dependent on the 
jet thrust axis, it is not dependent on the jet energy scale.

Figure 1 shows the $j_{T}$ distributions from jets in p+p collisions
with observed $E_T > 5$ GeV, for both the Monte Carlo and the min bias data. 
For hadron
$p_{T}>2.0$ GeV/c, the data and the Monte Carlo agree within 4\%.
 For charged hadrons
with $p_{T}>2.0$ GeV/c, $\left<j_{T}\right> =
515\pm50_{sys}$ MeV/c.

\section{Di-jet Studies}
\vspace*{-0.25in}
\begin{figure}[h]
\begin{spacing}{0.8}
\begin{flushleft}\hfill{\includegraphics*[%
  width=0.50\columnwidth]{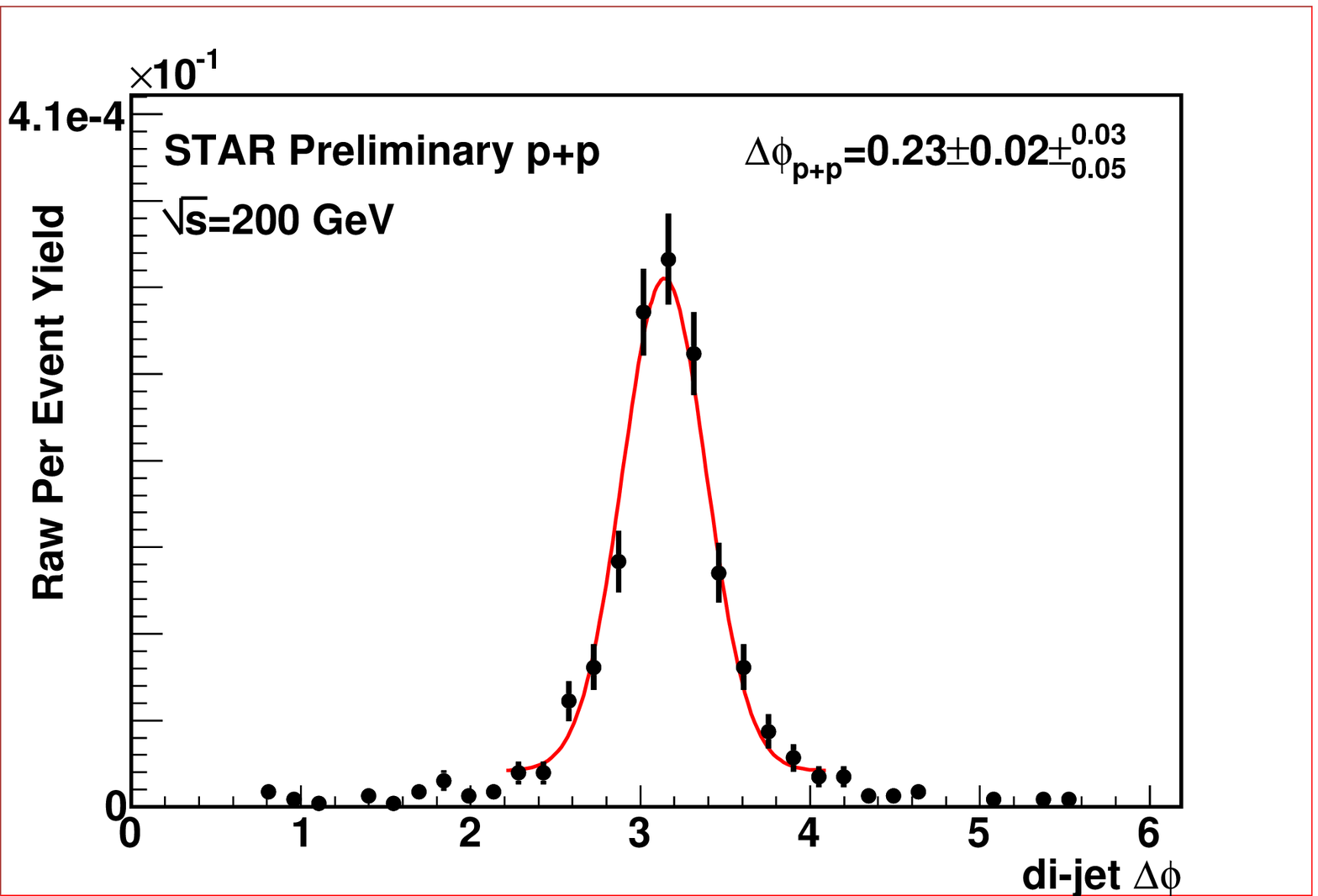}}{\includegraphics*[%
  width=0.50\columnwidth]{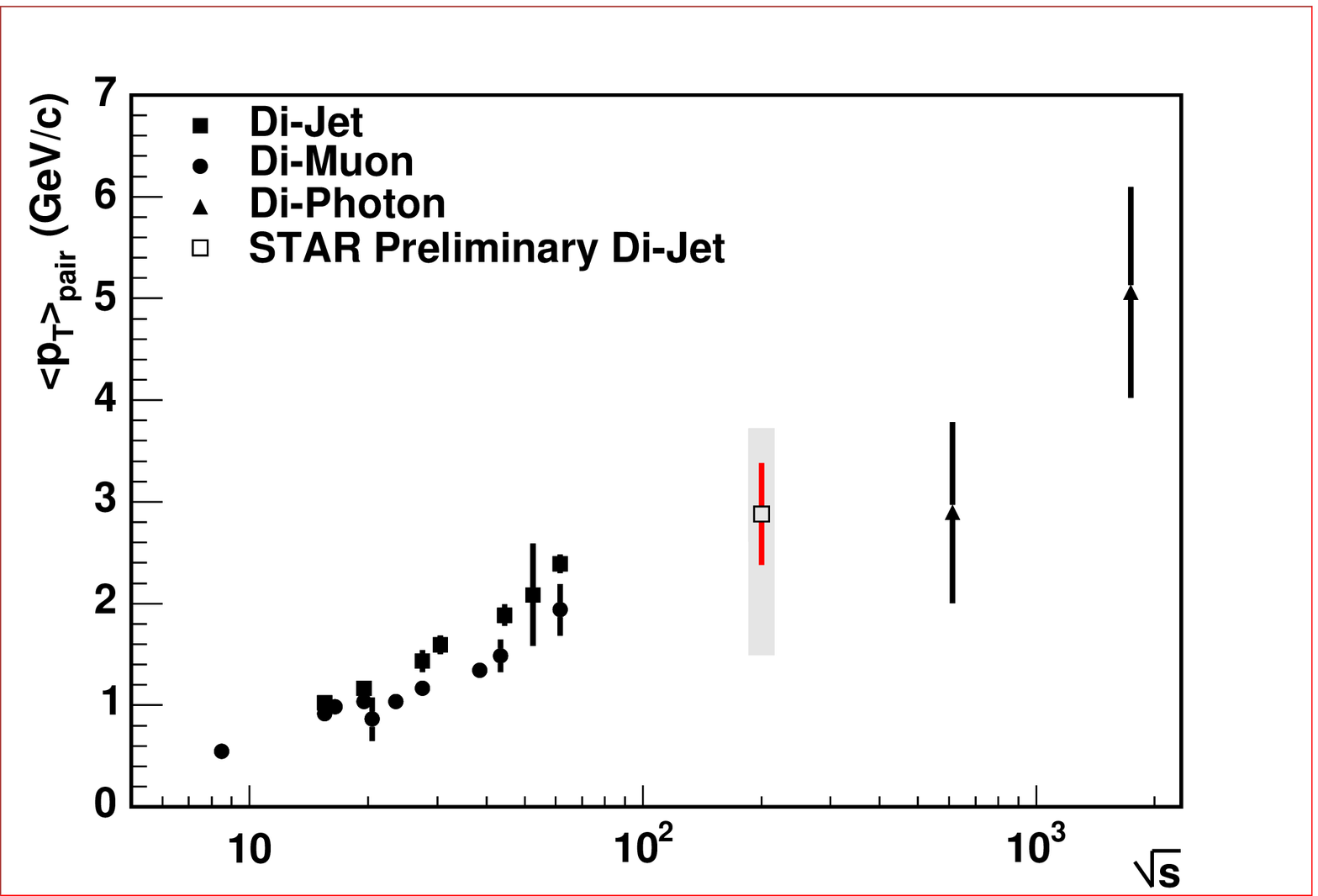}}\end{flushleft}
\vspace*{-0.25in}
\caption{The left panel shows the p+p di-jet $\Delta\phi$ distribution at
$\sqrt{s}=200$ GeV.  
The right panel shows comparison to the world data \cite{L. Apanasevich} for $\left<p_{T}\right>_{pair}$
as a function of $\sqrt{s}$.} 
\end{spacing}
\end{figure}
The di-jet opening angle is one of the basic observables obtained from di-jet 
events. For di-jets, $\Delta\phi=\pi$ in leading order QCD.
Gluon radiation broadens $\Delta\phi$, which
is measured by the per-parton root mean square 
$k_{T}\equiv
\sqrt{\left<k_{T}^{2}\right>}=\left<E_{T}\right>\sin\sigma_{\Delta\phi}$.
When reconstructing di-jets, one {}``trigger'' jet
with observed $E_T > 7$ GeV is reconstructed from both neutral and charged 
hadrons using the high tower which triggered the event. This establishes the 
di-jet energy scale, but also restricts the jet to positive $\eta$, due to  
the barrel EMC acceptance.
The other {}``away'' jet is reconstructed using charged particles only,
allowing the {}``away'' jet to range $-0.5<\eta<0.5$.  
This mixes both symmetric and asymmetric partonic collisions into the sample, 
just as in high-$p_T$ inclusive yields.

The left panel of figure 2 shows the di-jet $\Delta\phi$ distribution for p+p collisions. 
For these reconstructed p+p di-jets,
the corrected jet energy scale is $\left<E_{T}\right>=13.0\pm0.7_{sys}$ GeV. Using the
jet energy scale and the overall detector resolution obtained from
Pythia simulations, the intrinsic 
$\sqrt{\left<k_{T}^{2}\right>}=2.3\pm0.4\pm_{1.11}^{0.67}$ GeV/c.
The right panel of figure 2 shows the STAR p+p di-jet 
$\left<p_T\right>_{pair}
=\sqrt{\frac{\pi}{2}\left<k_{T}^{2}\right>}$ compared with
world data \cite{L. Apanasevich}.  
The left panel of Figure 3 shows the di-jet $\Delta\phi$ distribution for d+Au
collisions.
Taking
$k_{T_{obs}}^{2}=k_{T_{pp}}^{2}+k_{T_{nucl}}^{2}$,
we find the $\sqrt{\left<k_{T_{nucl}}^{2}\right>}=2.8\pm1.2\pm1.0$
GeV/c in d+Au collisions. The major systematic uncertainties 
are the jet energy scale in p+p and d+Au, the 
detector resolution, and treatment of the background in the 
di-jet $\Delta\phi$ distributions.  The present uncertainties are 
conservative estimates, 
and the capability exists at STAR for future improvement
of all the $k_{T}$ measurements.

A complementary di-hadron analysis at STAR correlates
high $E_T$ photons measured with the EMC and charged particles found in the 
TPC.  Photons provide significant advantages 
since the high tower trigger selects the events of interest.
The right panel of Figure 3 shows the results for d+Au collisions with two
different thresholds for the trigger photon $E_T$.
The di-hadron correlation strength increases and the width
narrows as the trigger photon $E_T$ is increased.  The increased $E_{T}$
also permits higher associated particle thresholds, which reduces the 
background.  When applied to Au+Au collisions, this technique
will provide much cleaner separation of jet-like correlations from anisotropic 
flow ($v_2$) compared to previous studies
with charged trigger particles \cite{btob}.
The di-hadron analysis finds that preliminary values 
for $\left<j_{T}\right> = 450\pm40\pm150$ MeV/c
and $\sqrt{\langle k_T^2 \rangle} = 
(1.9\pm0.2\pm0.3)/\left<z\right>$ GeV/c, where $\left<z\right>$, 
the average fragmentation fraction
of the trigger photon, is estimated to be the range $0.6 \sim 0.8$.
Within present systematic uncertainties, the jet analysis results and the 
di-hadron results agree.  For additional details, see \cite{subhasis}.

\begin{figure}[h]
\begin{spacing}{0.8}
\begin{flushleft}{\includegraphics*[%
  height=2.3in,
  width=0.48\columnwidth]{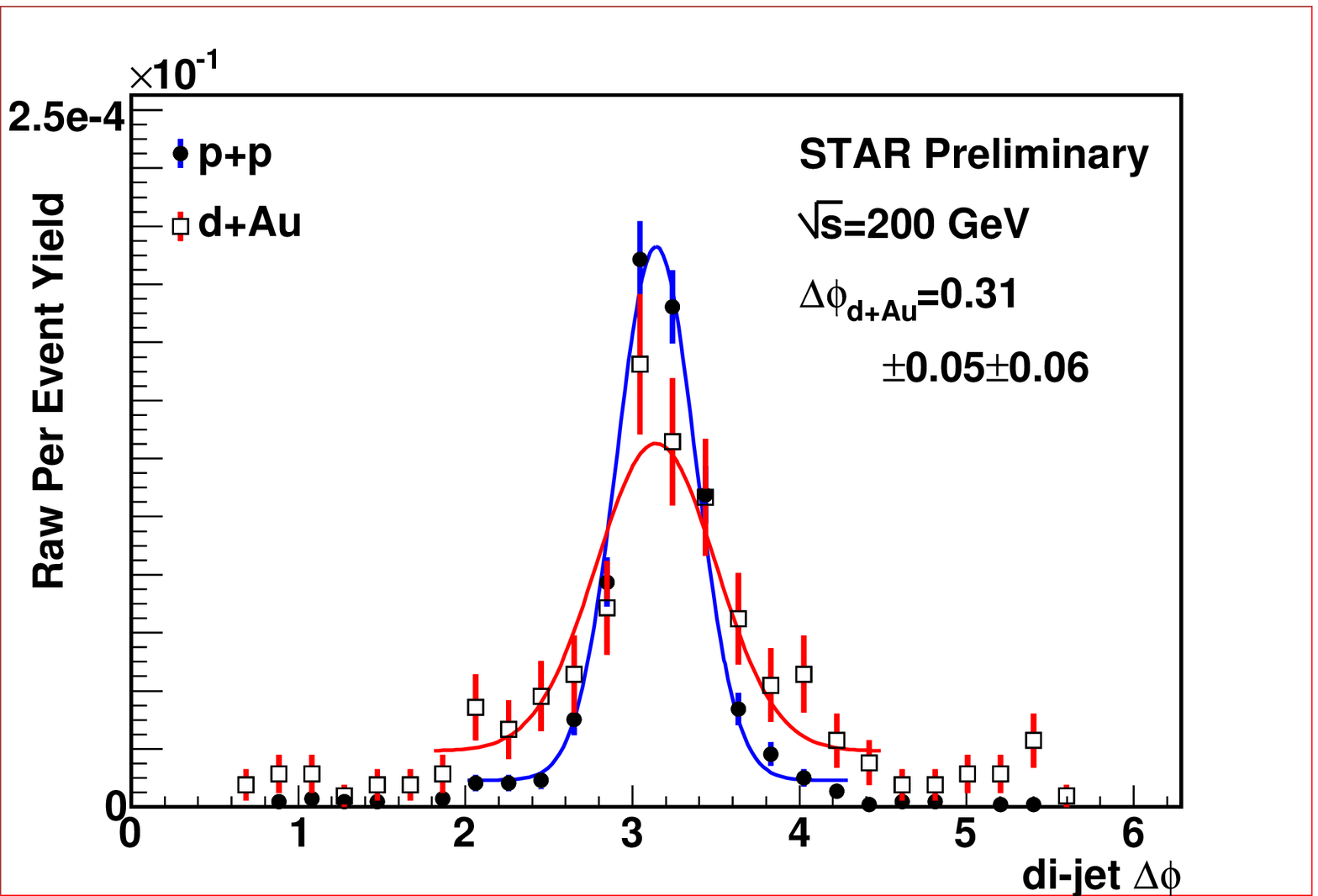}}\hfill{\includegraphics*[%
  height=2.3in,
  width=0.48\columnwidth]{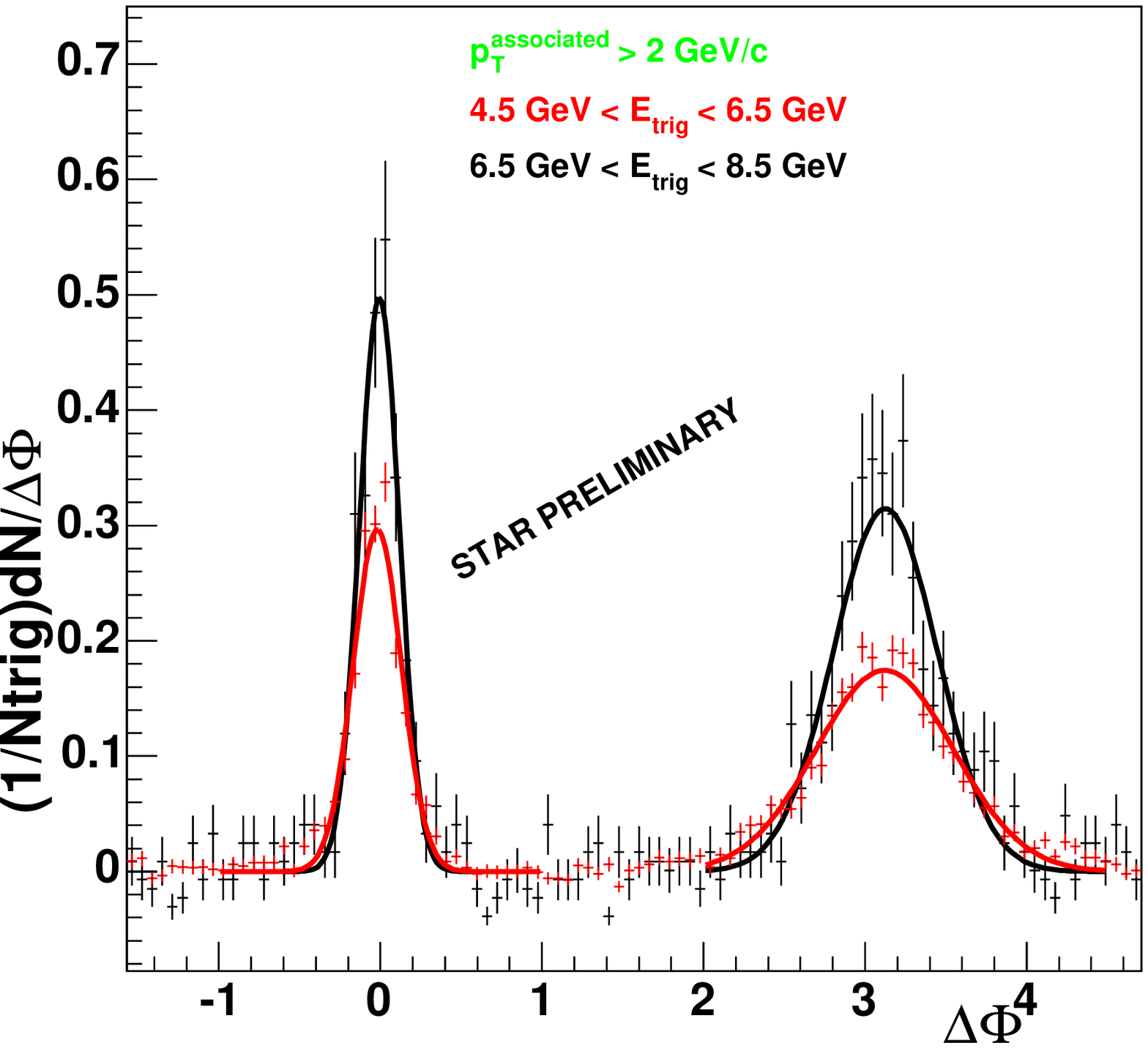}}\end{flushleft}
\vspace*{-0.25in}
\caption{The left panel shows the d+Au di-jet $\Delta\phi$ distribution at
$\sqrt{s}=200$ GeV.  $\Delta\phi_{d+Au}=0.31\pm0.05\pm0.06$. 
The right panel shows d+Au di-hadron $\Delta\phi$ from \cite{subhasis}.} 
\end{spacing}
\end{figure}
\vspace*{-0.25in}

\section{Conclusion}
Jets have been reconstructed in both p+p and
d+Au collisions. The p+p jets
have been used to measure both $j_{T}$ and $k_{T}$.
The $j_{T}$ 
results agree with the Monte Carlo simulation studies.
The d+Au $k_{T_{nucl}}$ value, though having large uncertainty, is in line 
with the expectation from previous experiments \cite{E609 Collaboration} 
indicating that $k_{T_{nucl}}$ is substantial compared with $k_{T_{pp}}$.
The $\langle j_T \rangle$ and $\sqrt{\langle k_T^2 \rangle}$ results from
full jet reconstruction agree well with similar measurements using
di-hadron correlations.
In addition to reducing uncertainties, future analysis will
focus on jet and di-jet cross sections in p+p and d+Au
collisions, as well as $R_{dAu}$ for jets (partons).

\end{document}